\documentclass[twocolumn,pra,aps,showpacs]{revtex4}

\usepackage{amsmath}
\usepackage{amssymb}
\usepackage[dvips]{graphicx}
\usepackage{psfrag}
\usepackage{pstricks}
\usepackage{pst-node}
\usepackage{pst-plot}

\newcommand{\qed}{\hspace*{\fill}$\square$}

 \newcommand{\C}{\mathbf{C}}

 \newcommand{\set}[2]{ \{\,#1\,|\,#2\,\}}

 \newcommand{\inver}[1]{\bar #1}

 \newcommand{\prima}{^\prime}

 \newcommand{\nin}{ \not\in}

 \newcommand{\ket}[1]{|#1\rangle}
 \newcommand{\bra}[1]{\langle #1|}
 
 \newcommand{\ketbradif}[2]{\ket{#1}\bra{#2}}
 \newcommand{\ketbra}[1]{\ketbradif {#1}{#1}}

 \newcommand{\GS}{\ket{\mathrm{GS}}}
 
 \newcommand{\Hilb} {\mathcal H}

 \newcommand{\Norm}[1] {\mathbf{N}_{#1}}

 \newcommand{\xpctd}[1]{\langle#1\rangle}

\begin{document}

\title[Short Title]{
Nested Topological Order}

\author{H. Bombin and M.A. Martin-Delgado}
\affiliation{
Departamento de F\'{\i}sica Te\'orica I, Universidad Complutense,
28040. Madrid, Spain.
}

\begin{abstract}
We introduce the concept of nested topological order in a class of
exact quantum lattice Hamiltonian models with non-abelian discrete
gauge symmetry. The topological order present in the models can be
partially destroyed by introducing a gauge symmetry reduction
mechanism. When symmetry is reduced in several islands only, this
imposes boundary conditions to the rest of the system giving rise to
topological ground state degeneracy. This degeneracy is related to
the existence of topological fluxes in between islands or,
alternatively, hidden charges at islands. Additionally, island
deformations give rise to an extension of topological quantum
computation beyond quasiparticles.
\end{abstract}

\pacs{71.10.-w, 11.15.-q, 03.67.Pp,  71.27.+a}

\maketitle

The concept of topological orders \cite{wenbook04} offers the
possibility of finding new states of matter with a common picture of
string-net condensation \cite{levinwen05} and other variants thereof
\cite{topo3D}. They correspond to examples of long range
entanglement in quantum many-body systems where those correlations
emerge in quantum states that are encoded in non-local degrees of
freedom of topologically ordered systems. Their global properties
are the source for yet another application as the suitable systems
to implement topological quantum computation
\cite{kitaev,freedman_etal00a,freedman_etal00b,rmp_topo_07}, a form
of fault-tolerant quantum computation intrinsically resistant to the
debilitating effects of local noise. Quantum field theories with an
spontaneous symmetry breaking mechanism of a continuous gauge group
down to a discrete group have been proposed as a scenario for
realizing their physics \cite{Bais_80, Bais_81, Krauss_Wilczek_89,
Preskill_Krauss_90, BSS_02a, BSS_02b, Bais_Driel_Wild_92}.

In this paper we introduce the concept of nested topological order
in a class of quantum lattice Hamiltonians. Our starting point are
the family of Kitaev's models \cite{kitaev}, which are labeled by a
discrete gauge group. Such models can be modified \cite{NAKMCC}
introducing an explicit symmetry breaking mechanism. Our aim is to
study the effect of `nesting' subsystems with a reduced symmetry
inside systems with the complete gauge symmetry. We will consider a
topologically ordered system divided in two regions, say $A$ and
$C$, and show that it is possible to partially destroy the
topological order in region $C$ in such a way that this imposes
boundary conditions to the subsystem $A$. The system $C$ can take
the form of several islands, which is why we talk about `nested'
topological order. The boundary conditions induce a topological
ground state degeneracy which is due to the possible values of
certain fluxes in between islands. As we will see, the values of
these fluxes correspond to the types of domain walls that exist in
$C$. If we allow the region $C$ to be deformed, then islands can be
initialized, braided and fused, giving an interesting extension of
the ideas of topological quantum computation beyond quasiparticles.

The models that we consider are string-net condensates in a 2D
lattice \cite{wenbook04}, \cite{levinwen05}. The configurations of
the lattice are regarded as string-net states: a collection of
labeled strings meeting at branching points. A string-net is closed
if certain conditions hold at branching points and there are no
loose ends. The ground state is a superposition of all possible
deformations of such closed string-nets, and excited states
correspond to configurations with loose ends: quasiparticle
excitations appear at the ends of strings. Now, to such system
Hamiltonians we can add string tension terms, which penalize with a
higher energy those configurations with longer strings. As such
terms get more important with respect to the original ones, longer
strings become less relevant in the ground state and finally the
topological order is destroyed as excitations get confined.
Alternatively, we can add suitable terms so that only part of the
topological order is destroyed. This is in fact the case for the
Hamiltonians $H_G^{N,M}$ that we consider \eqref{Hamiltoniano_NM},
which are labeled with a discrete group $G$ and two subgroups
$N\subset M\subset G$, with $N$ abelian and normal in $G$. If $N=1$
and $M=G$, we have the original topologically ordered models with
gauge group $G$ considered by Kitaev \cite{kitaev}. Otherwise, the
gauge symmetry is reduced the quotient group $G\prima = M/N$. In
particular, if $N=M$ the topological order is completely destroyed.

\emph{ {\bf Topological phases.}} The systems of interest are constructed
from a two-dimensional orientable lattice, of arbitrary shape. At
every edge of the lattice we place a qudit, a $|G|$-dimensional
quantum system with Hilbert space $\Hilb\prima_G$ and a basis $\ket
g$ labeled with the elements of $G$. The Hamiltonians read as
follows\cite{NAKMCC}
\begin{equation}\label{Hamiltoniano_NM}
H_G^{N,M}:=-\sum_{v\in V}
  A^M_v - \sum_{f\in F} B^N_f - \sum_{e\in E} \left (T_e^M + L_e^N
  \right),
\end{equation}
where the sums run over the set of vertices $V$, faces $F$ and edges
$E$. Explicit expressions for the terms in \eqref{Hamiltoniano_NM}
will be given below, but before that, we will discuss their physical
content. First, all the terms are projectors and commute with each
other, so that the ground state is described by conditions of the form
$P\GS=\GS$ with $P$ either a vertex, face or edge
operator. Excitations are gapped and localized; they correspond to
violations of the previous conditions and so can be related to
vertices, faces and edges; they are regarded respectively as electric,
magnetic and domain wall excitations.

We first recall the case $H_G:=H_G^{1,G}$\cite{kitaev}. For
non-Abelian groups $G$, vertex and face excitations are interrelated
and the excitation types, labeled as $(R,C)$, are dyons: $C$, the
magnetic part, is a conjugacy class of $G$ and $R$, the electric
part, is an irrep of $\Norm C$, the group $\Norm C:=\set{g\in
G}{gr_C=r_Cg}$, where $r_C$ is some chosen element of $C$. These
charges have a topological nature: if there are several excited
spots in the system, far apart from each other, there exist certain
global degrees of freedom which cannot be accessed through local
operators.

In the general case $H_G^{N,M}$ there are two new phenomena,
quasiparticle condensation and the appearance of domain wall
excitations. The latter have an energy proportional to their length
and can be labeled by pairs $(R,T)$, with $T\in M\backslash G/M$ and
$R$ an induced representation in $M$ of an irrep of the group $\Norm
T:=\set{m\in M}{mr_TM=r_TM}$, where $r_T$ is some chosen element of
$T$. Thus there exists a flux related to domain walls, with values
$(R,T)$; it is conserved in the absence of quasiparticle excitation,
so that domain walls only can end at them. As for condensation, we
will comment upon it below.

\begin{figure} \psfrag{alpha}{$\alpha$} \psfrag{beta}{$\beta$}
\psfrag{t1}{$\tau_1$} \psfrag{t2}{$\tau_2$ } \psfrag{t3}{$\tau_3$}
\psfrag{t4}{$\tau_4$} \psfrag{rho}{$\rho$} \psfrag{sigma}{$\sigma$}
\psfrag{A}{$S$} \psfrag{B}{$T$}
\includegraphics[width=8.2 cm]{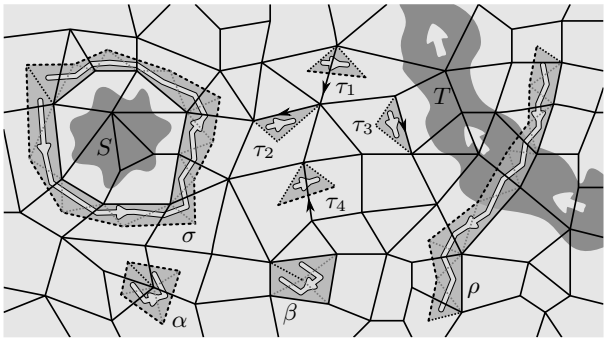}
\caption{Examples of lattice constructions. Although all the edges
must be oriented, only the orientation of some of them is shown. The
$\tau_i, i=1,2,3,4$ are triangles; the light thick arrow shows their
orientation. $\tau_1$ and $\tau_4$ are dual, the others are direct.
$\sigma$ is a closed ribbon; the projectors $K_\sigma^{R,C}$ give
the charge in the region $S$ that $\sigma$ encloses. $\rho$ is an
open ribbon; the projectors $J_\sigma^{R,T}$ give the domain wall
flux in the region $T$ in the direction of the arrows. $\alpha$ and
$\beta$ are minimal closed ribbons, enclosing respectively a single
vertex and face.}\label{Figura_red}
\end{figure}

\emph{{\bf Ribbon operators.}} In order to motivate the introduction
of ribbon operators, we first note that dyons, the excitations of
our system, are located at vertex-face pairs, which are called
sites. In Fig.~\ref{Figura_red} sites are represented as dotted
lines connecting the vertex to the center of the face. The basic
connectors between sites are triangles: just as an edge connects two
vertices, triangles connects two sites. A direct (dual) triangle
$\tau$ is composed by two sites and a direct (dual) edge $e_\tau$,
see Fig.~\ref{Figura_red}. Triangles can be concatenated to form
ribbons connecting distant sites. Ribbons are open if they connect
disjoint sites and closed if their ends coincide. The point is that
it is possible to attach to each ribbon $\rho$ certain operators
$F_\rho^{h,g}$, $h,g\in G$, which are very well suited to represent
excited states. For example, any state with only two dyons is a
linear combination of the states $F_\rho^{h,g}\GS$, with $\rho$ any
ribbon connecting the sites where the dyons are
located\cite{kitaev}. In fact, one can consider that ribbon
operators represent a process in which a particle-antiparticle is
created in one end of the ribbon and one of them is moved to the
other end.

In order to describe ribbon operators, we start with triangles,
which are the smallest ribbons. Recall that a triangle is formed by
two sites and one edge, direct or dual. Triangle operators act on
the qudit attached to that edge, and the action depends on the
orientation of the edge and the type of the triangle. The four
possible cases are illustrated in Fig.~\ref{Figura_red}. With the
notation of that figure, we have
$F_{\tau_1}^{h,g}=\delta_{g,1}\sum_k \ket{hk}\bra{k}$,
$F_{\tau_2}^{h,g}= \ketbra{g^{-1}}$, $F_{\tau_3}^{h,g}= \ketbra{g}$
and $F_{\tau_4}^{h,g}=\delta_{g,1}\sum_k \ket{kh^{-1}}\bra{k}$,
where the sums run over $G$. Then if $\rho$ is a ribbon formed by
the concatenation of the ribbons $\rho_1$ and $\rho_2$, we set
$F_\rho^{h,g}=\sum_k F_{\rho_1}^{h,k}F_{\rho_2}^{k^{-1}hk,k^{-1}g}$.
The terms in the Hamiltonians \eqref{Hamiltoniano_NM} are built from
ribbon operators. Let $F_\rho^{UV} :=|U|^{-1}\sum_{u\in U}\sum_{v\in
V} F_\rho^{u,v}$ for any subgroups $U,V\subset G$. Then
$A_v^M:=F_\alpha^{NG}$, $B_f^N:=F_\beta^{1N}$, $T_e^M:=F_\tau^{1M}$
and $L_e^N:=F_{\tau\prima}^{NG}$, with $\alpha$ and $\beta$ suitable
minimal closed ribbons as in Fig.~\ref{Figura_red} and $\tau$
($\tau\prima$) a direct (dual) triangle with $e=e_\tau$.

Ribbon operators commute with all the vertex operators $A_v^G$ and
face operators $B_f^1$, except with those at their ends. Moreover,
they can be characterized by this property\cite{NAKMCC}. This
suggests considering, for closed ribbons $\sigma$, those ribbon
operators which commute with all vertex and face operators, so that
they `forget' the single end of $\sigma$. It turns out that a linear
basis for such operators is given by a family of projectors
$K_\sigma^{R,C}$, labeled with the charge types $(R,C)$ of the
system $H_G$. In fact, if $\sigma$ is a boundary ribbon, that is, a
closed ribbon enclosing certain region $S$ as in
Fig.~\ref{Figura_red}, then $K_\sigma^{R,C}$ projects out those
states with total topological charge $(R,C)$ in $S$. As a result,
the ground state of $H_G$ can be described by the conditions
\begin{equation}\label{Descripcion_GS_Kitaev}
F^{G1}_\sigma \ket\psi = \ket\psi,
\end{equation}
which must hold for all boundary ribbon $\sigma$. This amounts to
impose that all disc shaped regions must have trivial charge because
$K^{e1}_\sigma=F^{G,1}_\sigma$, where $e$ is the identity
representation. In systems with Hamiltonian $H_G^{NM}$ we can use
the projectors $K_\sigma^{RC}$ to describe condensation. Namely, for
some charges \cite{NAKMCC} we have a ground state expectation value
$\xpctd{K_\sigma^{RC}}>0$ for any boundary ribbon $\sigma$, showing
that there exist a non-zero probability of finding such charges in a
given region.

Domain wall types can be obtained in a similar fashion in systems
with Hamiltonian $H_G^{NM}$. For any open ribbons $\rho$, those
ribbon operators that commute with all vertex operators $A_v^M$ and
face operators $B_f^N$ are linear combinations of certain projectors
$J_\rho^{R,T}$, with $(R,T)$ a domain wall type. If $\rho$ crosses
an area with domain wall excitations then $J_\rho^{R,T}$ projects
out those states with total domain wall flux $(R,T)$ across $\rho$.
For example, in Fig.~\ref{Figura_red} $\rho$ will measure the flux
of the excited region $T$ in the direction of the white arrows.

The ground states of \eqref{Hamiltoniano_NM} can also be described
in terms of conditions for ribbon operators, in particular by
\begin{equation}\label{Descripcion_GS_NM}
F^{MN}_\sigma \GS = \GS, \quad F^{NM}_\rho \GS = \GS,
\end{equation}
where $\sigma$ and $\rho$ are arbitrary boundary and open ribbons,
respectively. The first condition is related to vertex and face
excitations, and the second to edge excitations.

\emph{{\bf Nested phases.}} We are now in position to discuss a more
complicated system. In particular, we want to consider a surface
divided in two regions of arbitrary shape, $A$ and $C$, plus a third
region $B$ which is just a thick boundary separating them, included
so that the Hamiltonian does not have to change abruptly from $A$ to
$C$. The idea is to have a local Hamiltonian such that conditions
\eqref{Descripcion_GS_Kitaev} are satisfied in $A$, conditions
\eqref{Descripcion_GS_NM} in $C$ and the conditions
\begin{equation}\label{Descripcion_GS_B}
F^{NN}_\sigma \GS = \GS,
\end{equation} with $\sigma$ an arbitrary
boundary ribbon, in the whole system. The last condition is needed
to ensure that domain wall flux is preserved through region $B$, a
key ingredient of our construction as we will see. The ground state
of the Hamiltonian $H_0 := -\sum_v A_v^N -\sum_f B_f^N$ is described
precisely by \eqref{Descripcion_GS_B}. In addition, $H_0$ commutes
with $H_G$, $H_G^{NM}$. Indeed, a Hamiltonian of the form
$H'=H_G+\lambda H_0$, $\lambda\geq 0$, only differs from $H_G$ in
the gap for some excitations, and the same is true for $H_G^{NM}$.
The Hamiltonian that we want to consider takes the form $H =
H_0+\lambda H_G+\mu H_G^{NM}$, where $\lambda,\mu \geq 0$ vary
spatially so that $\lambda=1$ and $\mu=0$ in $A$ and $\lambda=0$ and
$\mu=1$ in $C$. If we take $\lambda\mu=0$, the ground state has the
desired properties but there exists some local degeneracy at $B$.
This local degeneracy can be lifted if $\lambda$ and $\mu$ are
allowed to overlap, but on the other hand if the overlap is too big,
it could produce a level crossing taking the ground state of $H$ out
of that of $H_0$, which spoils conditions \eqref{Descripcion_GS_B}.

\begin{figure} \psfrag{a}{a} \psfrag{b}{b} \psfrag{c}{c}
\psfrag{r1}{$\rho_1$} \psfrag{r2}{$\rho_3$} \psfrag{r3}{$\rho_4$}
\psfrag{r4}{$\rho_2$ } \psfrag{r5}{$\rho_5$} \psfrag{r6}{$\rho_6$}
\psfrag{r7}{$\rho_7$} \psfrag{r8}{$\rho_8$}
\includegraphics[width=8.2 cm]{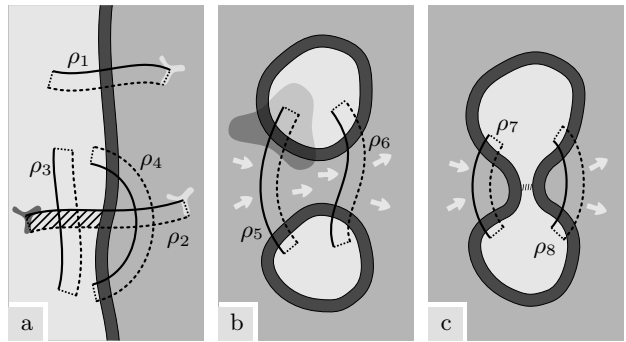}
\caption{ In this figure regions $A$, $B$ and $C$ are shaded
respectively with medium, dark and light gray. Ribbons $\rho_i,
i=1,\ldots, 8$ are displayed as pairs of solid and dashed parallel
lines which correspond respectively to their direct and dual edges.
Light spots at the end of ribbons represent excitations in $A$ and
the dark one an excitation in $C$. The striped areas are domain wall
excitations. (a) Due to condensation, suitable ribbon operators
attached to $\rho_1$ will create an excitation in $A$ but no
excitation in $C$. Ribbon operators attached to $\rho_2$ can create
a domain wall excitation in $C$. The resulting state $\psi$ is such
that $J_{\rho_3}^{R,T}\ket\psi=J_{\rho_4}^{R,T}\ket\psi$. (b) Both
$\rho_5$ and $\rho_6$ measure the flux in between the islands. If
$O$ is an operator with support in the shaded area an takes ground
states to ground states, it cannot change the flux. (c) If the
previous islands are deformed till they fuse, the flux measured by
$\rho_7$, $\rho_8$ will remain the same as it was for $\rho_5$,
$\rho_6$. If it is nontrivial, opposite border charges are present
at the sides of the meeting point. }\label{Figura_holes}
\end{figure}

\noindent {\em  {\bf Quasiparticle dilution.}} Our aim is to
understand the effects of the nested region $C$ on the topologically
ordered region $A$. A first effect is the possibility to locally
create or destroy single quasiparticle excitations in the vicinity
of the $A$-$C$ border, something prohibited in systems with
Hamiltonian $H_G$ due to charge conservation. In terms of ribbon
operators, this is reflected in the fact that for any $\rho_1$
connecting $C$ to $A$, as the one in Fig.~\ref{Figura_holes}(a), a
state of the form $\sum_{m\in M} F_{\rho_1}^{mn\inver m,mg}\GS$,
$n\in N$, contains no excitation at $C$. In terms of quasiparticle
processes, this corresponds to create a particle-antiparticle pair
in $A$ and then move one of them into $C$, where it disappears
because it is condensed.

\noindent {\em {\bf Domain wall dilution.}} A second effect is
related to the existence of domain walls in region $C$. Consider
again a ribbon $\rho_2$ connecting $C$ to $A$, see
Fig.~\ref{Figura_holes}(a). Some of the states of the form $\ket\psi
= \sum_{h,g} c_{h,g} F^{h,g}\GS$, $c_{h,g}\in \C$, will contain edge
excitations all along the portion of $\rho_2$ contained in $C$, for
example those with $c_{h,g}\neq 0$ for some $g\in G,h\nin M$. These
excitations form a domain wall, to which we can relate a type or
flux given by the projector $J_{\rho_3}^{RC}$, where $\rho_3$ is a
ribbon that lies in $C$ and crosses the domain wall, see
Fig.~\ref{Figura_holes}(a). Such a ribbon can be deformed without
crossing any quasiparticle excitation onto another ribbon $\rho_4$
that only has its endpoints in $C$ and thus avoids the domain wall,
so that $J_{\rho_3}^{R,T}\ket\psi=J_{\rho_4}^{R,T}\ket\psi$ due to
\eqref{Descripcion_GS_B}. Both ribbon operators are measuring the
same domain wall flux. However, in the case of $\rho_4$ the flux is
being measured in $A$, where the domain wall gets diluted as it
turns into a condensed string. Note that $J_{RC}^{\rho_4}$ cannot
detect changes in the interior of $C$. In this regard, if we
restrict our attention to region $A$, domain wall flux projectors
from ribbons like $\rho_4$, that is, which enclose a portion of the
$A-C$ border, can be related to charges $(R,T)$ that lie in that
piece of the $A-C$ border.

\noindent {\em {\bf Induced topological fluxes.}} Things get even
more interesting if we consider that $C$ consists of several
disjoint parts. For example, consider a plane and choose as the
region $C$ two islands $C_1$ and $C_2$, see
Fig.~\ref{Figura_holes}(b). Now instead of considering a domain wall
flux coming out from a region of $B$ (such as the one measured by
$\rho_4$ in Fig. ~\ref{Figura_holes}(a)), we consider the flux in
between the two islands (as indicated by the arrows in
Fig.~\ref{Figura_holes}(b)). This is the flux measured by the
projectors $J_{\rho_5}^{R,T}$, where $\rho_5$ is any ribbon that
connects the islands, as in Fig.~\ref{Figura_holes}(b). The point is
that such a flux is a global (topological) property as long as the
islands are distant. Indeed, measuring the flux requires an operator
with a support connecting $C_1$ and $C_2$. And, if an operator
changes the flux, its support must loop around $C_1$ (or $C_2$).
Suppose to the contrary that $O$ is an operator that leaves the
ground state invariant and has a support not enclosing $C_1$, as the
shaded region in Fig.~\ref{Figura_holes}(b). Let $\rho_6$ be another
ribbon connecting the islands but lying outside the support of $O$.
Due to \eqref{Descripcion_GS_B} we have $J_{\rho_5}^{R,T}\GS =
J_{\rho_6}^{R,T}\GS$, so that $[J_{\rho_5}^{R,T},O]\GS=
[J_{\rho_6}^{R,T},O]\GS=0$ and thus $O$ does not change the flux.
Those operators which do change the flux are related to processes in
which a particle-antiparticle pair is created, one of them loops
around $C_1$ and they meet again to fuse into a charge that
disappears into $C_1$.

\noindent {\em {\bf Topologically protected subsystems.}} It follows
that there exist a topological degeneracy in the ground state,
related to the distinct values that the flux in between $C_1$ and
$C_2$ can take. For example, if $N=M=1$ the flux can take any value
$g\in G$. In general, for a $C$ composed of multiple disconnected
regions, the degeneracy of the ground state depends on $N$, $M$ and
the topology of $A$. Now, it is natural to ask how does this
protected space compares with the one due to to the existence of
several separated excitations in $A$. In other words, do islands add
something new? This can be positively answered through an example:
two excitations give no protected subspace \cite{kitaev}, but we
have just seen the contrary for the case of two islands. Perhaps
more dramatically, for abelian groups $G$ the protected subsystem is
always trivial whatever the amount of excitations, but this is not
the case for islands. Nevertheless, islands can be compared to
excitations, in the following sense. An island can hold certain
charge values, which can be measured using ribbon operators that
enclose the island, as in the case of an excitation. The difference
between a charged island and a charged excitation is that the local
degrees of freedom of the excitation become global in the case of
the island: this is the origin of the additional dimensionality of
the protected subsystem.

\noindent {\em {\bf Braiding.}} The physics of the system so far has
a static nature. If we want to consider the setting as an scenario
for quantum computation, then the possibility of dynamically
deforming the region $C$ must be included in it. Such deformations
need not be strictly adiabatic, but the state should be kept in the
subspace defined by conditions
(\ref{Descripcion_GS_Kitaev}-\ref{Descripcion_GS_B}) at all time. We
can then braid islands to perform unitary operations, in complete
analogy with quasiparticle braiding. It is also natural to enrich
the physics by considering islands with different $(N,M)$ labels,
increasing the variety of protected subsystems.

\noindent {\em {\bf Fusion.}} We must consider also the analogue of
the quasiparticle fusion processes, which is the way in which
measurements are carried out in topological quantum computation.
There are two natural ways in which global degrees of freedom can be
made local. The first is to decrease the size of an island till it
dissapears leaving a small charged region. The outcome of such a
process is the charge, which can be measured but not changed
locally. The second way is closer to the idea of fusion. Indeed, it
is also a fusion, but of islands instead of quasiparticles. The idea
is depicted in Fig.~\ref{Figura_holes}(c). As two islands of the
same $(N,M)$ type get closer, some of the ribbon operators
connecting them become small and thus the flux between the islands
is exposed to local measurements. If we continue the approach till
the islands meet, the flux will take the form of a domain wall
excitation at the meeting place, as in Fig.~\ref{Figura_holes}(c).
Due to confinement the domain wall can decay to several smaller
walls, but there is something that will not disappear, the two
border charges in its ends on $B$. As explained in the caption of
Fig.~\ref{Figura_holes}(c), the appearance of this border charges
can be seen directly in terms of ribbon operators. Regarding the
initialization of the system, reverse processes can be used. That
is, if an island is divided in two, the topological flux in between
them will be trivial, and if an island is created from the vacuum,
it will have trivial charge. In both cases the reason is that
topological properties cannot be changed by local processes.

\noindent {\em Acknowledgements} We acknowledge financial support
from a PFI grant of the EJ-GV (H.B.), DGS grants under contracts BFM
2003-05316-C02-01, FIS2006-04885 (H.B., M.A.M.D,), and the ESF
Science Programme INSTANS 2005-2010 (M.A.M.D.).

\end{document}